# Enhancing HPC Security with a User-Based Firewall

Andrew Prout, William Arcand, David Bestor, Bill Bergeron, Chansup Byun, Vijay Gadepally, Matthew Hubbell,
Michael Houle, Michael Jones, Peter Michaleas, Lauren Milechin, Julie Mullen, Antonio Rosa, Siddharth Samsi,
Albert Reuther, Jeremy Kepner

MIT Lincoln Laboratory, Lexington, MA, U.S.A.

*Abstract*—High Performance Computing (HPC) systems traditionally allow their users unrestricted use of their internal network. While this network is normally controlled enough to guarantee privacy without the need for encryption, it does not provide a method to authenticate peer connections. Protocols built upon this internal network, such as those used in MPI, Lustre, Hadoop, or Accumulo, must provide their own authentication at the application layer. Many methods have been employed to perform this authentication, such as operating system privileged ports, Kerberos, munge, TLS, and PKI certificates. However, support for all of these methods requires the HPC application developer to include support and the user to configure and enable these services. The user-based firewall capability we have prototyped enables a set of rules governing connections across the HPC internal network to be put into place using Linux netfilter. By using an operating system-level capability, the system is not reliant on any developer or user actions to enable security. The rules we have chosen and implemented are crafted to not impact the vast majority of users and be completely invisible to them. Additionally, we have measured the performance impact of this system under various workloads.

## I. INTRODUCTION

The merger of traditional HPC systems, big data clouds, and elastic computing clouds has highlighted new challenges in managing a combined SuperCloud. One key challenge is addressing the security concerns that mixing this wide variety of applications brings to the system. The ability for applications to operate among the many nodes of the system implicitly requires them to make use of the network in some manner. However, on HPC systems the responsibility of ensuring the network is used securely has traditionally been delegated to each end-user's application.

Few applications are prepared to operate in this multiuser environment out of the box. Configuring the appropriate security controls in an application, assuming it has the options for security controls at all, or API layers, such as MPISec I/O [Prabhakar 2009], is often regulated to the "Advanced Configuration" section of manuals, and the topic is skipped in many how-to guides. Fewer solutions still are prepared to address the sharing of each distributed node among jobs of different users, each trying to use the network to communicate with their own distributed peers [Pourzandi 2005].

Even with services to assist users in standing up authenticated network connections, such as preprovisioned PKI certificates [Hunt 2001] in a virtual smartcard [Prout 2012] for use with TLS [Dierks 2008], bugs, poor documentation, and lack of user motivation prevent effective implementation of basic security measures. Leaving the security to end-users' applications is also particularly problematic when the application is still a work-in-progress. Developers cannot be expected to have a fully featured and tested security subsystem guaranteed to be ready and implemented the first time they perform a test run of their application. They must be given a development environment with enough security external to their application to allow for development and testing.

Considering traditional system or network-level firewalling techniques to address these challenges immediately runs into problems. The HPC scheduler can assign applications to run on any node in the system, different users can be sharing the same node, and users can utilize any unprivileged port number for their applications. Multiple users sharing the same node require localhost-based connections to be controlled as well; these connections are necessarily out of scope of even dynamic firewall solutions that focus on perimeter security concerns [Wiebelitz 2009, Green 2004].

We have addressed these challenges by building a user-based firewall into the HPC. It operates on the system level for all applications without any modification and with a low overhead to the network speed. The decisions made by this firewall are completely based on the user and group information of the processes on each end of the network communication and not the node they run on.

The organization of the rest of this paper is as follows. Section II describes the operational model and user experience of this system. Section III describes the technologies used to create the user-based firewall: Linux Netfilter, Ident2, NetID, and the MIT SuperCloud system, which was used to conduct the performance measurements. Section IV shows the performance results and overhead of the system. Section V describes future work in this area. Section VI summarizes the results.

## II. USER-BASED FIREWALL OPERATIONAL MODEL

The primary goal of the MIT SuperCloud user-based firewall was to create a separation between network-enabled processes of one user and those of all other users, while ensuring that legitimate intentional network traffic is allowed to pass. We started with the assumptions that the system implements the user private group model and that only system services are configured as exempt. The goal of our user-based

This material is based upon work supported by the National Science Foundation under Grant No. DMS-1312831. Any opinions, findings, and conclusions or recommendations expressed in this material are those of the author(s) and do not necessarily reflect the views of the National Science Foundation.

firewall configuration is to allow connections if they meet one of the following rules:

- connector process user matches the listener process user; or
- connector process primary group or any of the supplemental groups match the listener process primary group; or
- listening port is privileged (less than 1024); or
- connector or listener process user is configured as exempt.

With the rules above, we have ensured that any network services started by a user will by default be accessible only to that user. However, if a user is working collaboratively and wishes to intentionally allow other users to access network services they start, he or she can change the primary group of their working session from their private group to the appropriate project group. This change can be accomplished by using the `newgrp` or `sg` command and is preserved across job launches with both the Grid Engine and SLURM schedulers.

### III. TECHNOLOGIES

#### A. MIT SuperCloud

The MIT SuperCloud software stack enables traditional enterprise computing and cloud computing workloads to be run on an HPC cluster [Reuther 2013, Prout 2015]. The software stack runs on many different HPC clusters based on a variety of hardware technologies. It supports systems with both 10 GigE and FDR InfiniBand running Internet Protocol over InfiniBand [RFC 4391].

The MIT SuperCloud software stack, which contains all the system and application software, resides on every node. Hosting the application software on each node accelerates the launch of large applications (such as databases) and minimizes their dependency on the central storage.

#### B. Netfilter

Netfilter is the packet-filtering framework of the Linux kernel. It is often known to system administrators by the name of the userspace program that configures it: iptables. Using this framework allows firewall rules to be implemented on a Linux-based system for packets entering or exiting the system, or even passing between programs within a system over the loopback adapter. The desired behavior is achieved by loading and configuring together various netfilter modules. Two modules, conntrack and queue, are critical to the operation of the user-based firewall.

The netfilter conntrack module is used to track active connections. This module allows the netfilter to be a stateful firewall for Linux. By tracking a connection through its lifecycle, conntrack enables a simple way to permit follow-up packets on an existing connection to pass without a specifically crafted rule to match those packets in isolation. By applying a simpler test to follow-up packets, conntrack allows more complex and resource intensive rules to be applied to the initial connection establishment without affecting long-term throughput.

The netfilter queue module is used to dispatch packets to a user-space application for evaluation. This application can then issue a verdict to drop, pass, or modify the packet. The application can perform whatever additional processing it wishes prior to issuing a verdict on a packet. The kernel does limit how many packets can be waiting in the queue for a verdict, by default to 1024, so the application must ensure a timely verdict is issued on each packet to flush the queue. If the queue reaches its limit, the default behavior is to drop all new packets.

#### C. Ident2

One of the key building blocks in enabling a user-based firewall is to identify the user of each end of a network connection. This problem is not new, and has previously been solved by the ident protocol [RFC 1413] and its predecessor Authentication Service [RFC 912]. However the original protocol has certain drawbacks: it only works on TCP sockets, it is widely acknowledged to be a security risk because it leaks information about users of a system, and it only returned the basic username of the user.

We developed a new daemon and protocol that we call ident2 that seeks to solve these drawbacks. Our ident2 daemon running on each system provides a Unix domain socket to which applications submit queries for either local or remote network connections. By using a Unix domain socket, we can use standard filesystem permissions to control who can access the ident2 service and the information it provides. Our new ident2 protocol operates on UDP instead of TCP and requires a privileged port number to be used, ensuring our ident2 daemons only speak to their peer ident2 daemons. Along with IP range restrictions, these restrictions limit the ability to query our service and solve the unrestricted information leakage.

Once a query for an IP and port pair has been received, either for a local connection or from a peer ident2 daemon, ident2 will query the kernel to find the associated socket number for the network connection. For TCP, this step is done via the Netlink NETLINK_INET_DIAG interface with the inet_diag_req structure. For queries regarding UDP sockets, the output of /proc/net/udp is parsed instead, as inet_diag_req does not support UDP sockets.

Mapping a socket number to detailed user information is not straightforward. While network sockets do have a uid field, this is not the detailed information we are looking for. For that we must map a socket back to its owning process. This mapping can be accomplished by searching the list of open file handles for the socket. This mapping is not deterministic however, as more than one process can have the same socket open simultaneously. This indeterminism does not significantly affect our use case, as it would be unusual for multiple processes owned by different users to have open file descriptors to the same socket.

Once an associated process for the socket is found, that process's identity is queried and used to craft the reply, which includes the effective user that owns the process, the effective primary group, supplemental groups, and process identifier. This reply is then forwarded back to the peer ident2, if necessary, and then back to the querier via the Unix domain socket.

*D. NetID*

The central component of our user-based firewall is the netid daemon. It acts as a netfilter queue endpoint, receiving packets from the netfilter queue and issuing verdicts as to whether those packets should be passed or dropped. Upon receiving a packet from the queue, the service will query ident2 for both the local and remote user information. Upon receiving a reply from either query, it will make a preliminary check to determine if the connection can be allowed based on the one response. A connection could be allowed based on only one response if the connecting or listening user is configured to be exempt from the user-based firewall rules, which is done for many system services. Otherwise, the daemon will continue to wait until both queries are answered, or a predefined timeout is reached.

Once both queries are answered, the daemon will make a decision on the packet. If an accept verdict is issued on a packet, the connection completes normally. The netfilter conntrack module will add the connection to its tracking list, and the netfilter rule to send the packet to the queue will not trigger again for this connection. If the connection is denied, a drop verdict is issued and an ICMP destination unreachable notification is sent to the originating host. In the case of a TCP connection, this ICMP error message will also indicate to the source system that the automatic retry mechanism should stop attempting to connect. If the timeout is reached, a drop verdict is issued, but no ICMP notification is sent as retries could be appropriate.

## IV. PERFORMANCE

The performance of netfilter and conntrack have already been well explored [Hoffman 2003, Kadlecsik 2004]. By utilizing the existing netfilter conntrack module we are able to limit the impact of our user-based firewall to a delay in connection initiation.

We have benchmarked two metrics for performance of the user-based firewall: overall throughput and connection initiation delay. We performed these tests on an HPC system utilizing 10 GigE for its interconnection. To test the overall throughput, we transferred increasing amounts of data through a single TCP network connection using netcat. To test the connection initiation delay, we timed how long it took to create 1000 new TCP network connections and transfer one byte with an increasing number of threads.

As expected, the bulk data transfer showed an imperceptible difference between the measurements with and without the user-based firewall. The netfilter conntrack module, which is widely used and has been well optimized for exactly this use case, ensures that the user-based firewall only has to validate the first packet of the TCP connection.

The rapid connection use case showed a much greater impact. We saw approximately a 90% increase in total time taken to complete the test for 1 to 7 threads (1,000 to 7,000 total connections). When reaching 8 threads and higher, the impact began to rise significantly.

To mitigate the impact of this performance hit, we developed a library that could be integrated into a user's application by using the LD_PRELOAD environment variable to notify ident2 in advance of sockets created by that process. This precaching of socket to process mapping reduced the impact to approximately 35% on the 1 to 5 thread tests. At greater than 8 threads, this precaching is no longer beneficial in its current form. Further optimizations are likely possible to the cache handling and we may explore them in the future.

## V. FUTURE WORK

In the future, we will migrate ident2 from the inet_diag_req structure to the inet_diag_req_v2 structure and investigate better ways of handling UDP traffic. The inet_diag_req_v2 structure became available in the Linux kernel version 3.3. Our hope is that changing to the inet_diag_req_v2 structure will improve performance by allowing more precise connection

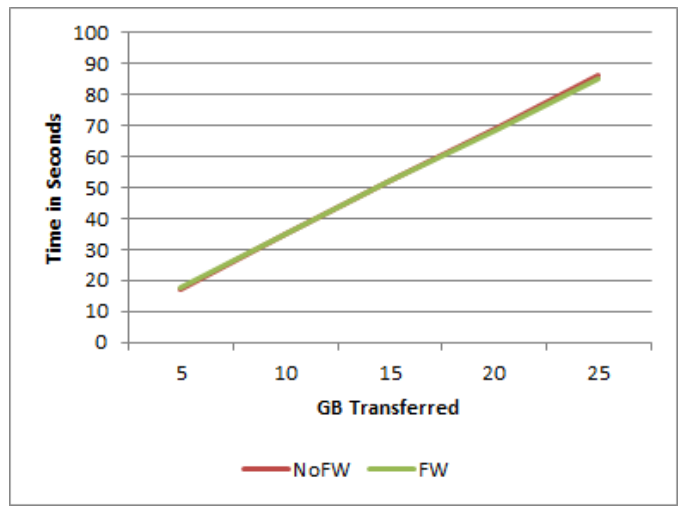

Figure 1. Time to complete transfers of various sizes by using a single-threaded netcat command with and without the user-based firewall enabled.

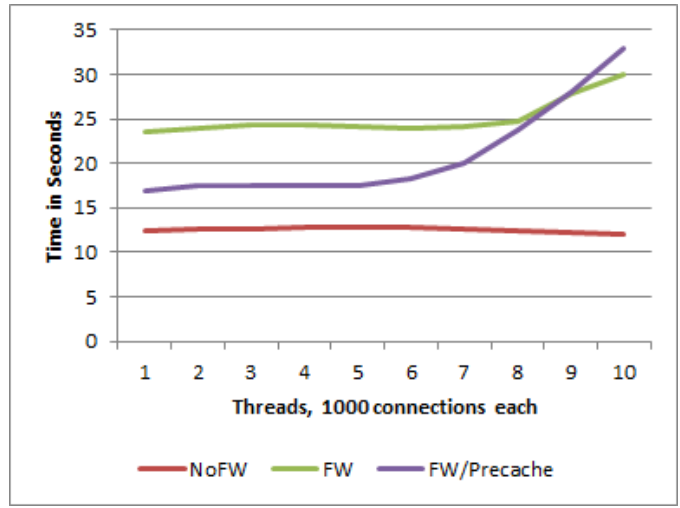

Figure 2. Time to complete 1000 new connections per thread and to transfer one byte each via netcat, without the user-based firewall, with the user-based firewall, and with the precache feature.

matching in the kernel and enabling us to query UDP information from Netlink instead of proc.

Future work will also investigate better ways of handling UDP traffic. Assigning "listener" and "connector" roles to UDP traffic to align with our operational model's rules is an imperfect fit. Additionally, "fire and forget" UDP programs can close their sockets or terminate before we query their information, leading to improper packet drops.

VI. SUMMARY

The user-based firewall we have implemented seamlessly applies the concept of user ownership to network traffic within an HPC cluster for the vast majority of workloads. The user-based firewall requires no changes to users' code to be implemented and minimal changes to their working patterns when they need to work collaboratively with other users or have a high rate of short connections.

The protections provided by our user-based firewall have already enabled our team to develop and deploy solutions that integrate third-party software faster by allowing us to avoid extensive effort that would have been required to integrate security directly into those tools. As the user-based firewall is based solely on process identity, it is agnostic to the user's initial authentication method to the system and is not reliant on a password-based authentication model. We have integrated ident2 support into our existing web portal efforts that support smartcard login [Reuther 2010]. By extending the same rules from the operational model to websocket forwarding through the portal, we have been able to rapidly enable new use cases for the system.